\documentclass[a4paper,10pt]{article}

\usepackage{a4wide}

\usepackage{amsmath}
\usepackage{amsfonts}
\usepackage{amssymb}

\usepackage{cite}

\usepackage{graphicx}

\usepackage{hyperref}

\usepackage{color}

\newcommand{\bm}{\mathbf}

\newcommand{\din}{\lrcorner}

\DeclareMathOperator{\extd}{d}
\DeclareMathOperator{\sgn}{sgn}

\title{Negative index of refraction, perfect lenses and transformation optics -- some words of caution.}
\author{Luzi Bergamin\thanks{Email: Luzi.Bergamin@tkk.fi}\\ \begin{small}Aalto University, School of Science and Technology                                                                                                              \end{small}\\ \begin{small} Department of Radio Science and Engineering\end{small}\\\begin{small}FI--00076 AALTO, Finland\end{small} \and Alberto Favaro\thanks{Alberto.Favaro04@imperial.ac.uk}\\
\begin{small}Department of Physics\end{small}\\
\begin{small}Imperial College London\end{small}\\
\begin{small}Prince Consort Road, SW7 2AZ, UK\end{small}}

\date{\today}

\begin{document}

\maketitle

\begin{abstract}
  In this paper we show that a negative index of refraction is not a direct implication of transformation optics with orientation-reversing diffeomorphisms. Rather a negative index appears due to a specific choice of sign freedom. Furthermore, we point out that the transformation designed lens, which relies on the concept of spacetime folding, does not amplify evanescent modes, in contrast to the Pendry-Veselago lens. Instead, evanescent modes at the image point are produced by a duplicated source and thus no imaging of the near field (perfect lensing) takes place.
\end{abstract}

\section{Introduction}
Negative index of refraction and perfect lenses \cite{Pendry:Pl2000,Veselago:1968Nr,Ramakrishna:2005Rp,Veselago:2006Nr} have become some of the most important concepts in metamaterials. The theoretical design of such devices was considerably stimulated by the claim \cite{Leonhardt:2006Nj} that a negative index of refraction could be understood from transformation optics as a transformation of space that inverts its orientation. Based on this idea, not only the flat perfect lens was re-interpreted as a folding of space \cite{Leonhardt:2006Nj}, but also lenses with different shapes \cite{Pendry:2003Jp,Pendry:2003Cl,Yan:2008Pr,Bergamin:2008Mm} were proposed. In many of these concepts a perfect lens was established by a folding of space, such that three points in laboratory space (one on each side of the lens and one inside the lens, as in Fig. 3 of \cite{Leonhardt:2006Nj}) corresponded to a single point in the virtual electromagnetic space (used to derive the medium properties). Based on these successes it was natural to conclude that transformation optics was an ideal tool to design perfect imaging devices. Recently, it was even suggested \cite{Leonhardt:2009Td,Leonhardt:2009Pi} that perfect imaging should be seen as the result of multi-valued maps rather than an effect of the amplification of evanescent waves. \emph{Our aim} is to critically review these ideas and to reassess the status of negative index of refraction and perfect lenses within transformation optics.

\begin{figure}[t]
 \centering
 \includegraphics[width=0.8\linewidth]{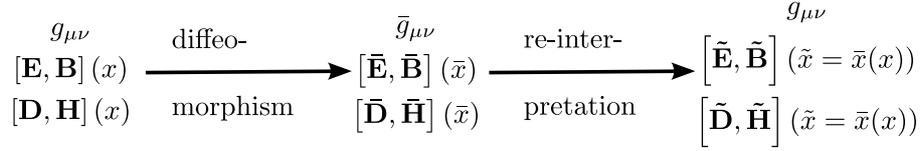}
\caption{Illustration and notation of transformation optics.}
\label{fig:TOnotation}
\end{figure}

Let us shortly review the basic steps of transformation optics and at the same time formulate the main questions to be addressed. Transformation optics intends to ``mimic'' a different spacetime by means of a special medium. In most cases this virtual spacetime (called electromagnetic space) is considered as diffeomorphic to the real spacetime (laboratory space) and transformation optics can also be seen to mimic a transformation of coordinates. In that case one starts by writing down a vacuum solution $\bm D = \varepsilon_0 \bm E$, $\bm B = \mu_0 \bm H$ of the Maxwell equations\footnote{Transformation optics relies on generic coordinates and thus an appropriate formalism has to be employed. Here, we follow Refs.\ \cite{Leonhardt:2006Nj,Leonhardt:2008Oe,Bergamin:2008Pa} and use component notation in conjunction with the Einstein summation convention. Thus in all equations a summation over repeated indices is assumed. Latin indices refer to space and the sum is performed over the values $i=1,2,3$. Greek indices are spacetime indices, the sum runs over $\mu=0,1,2,3$, whereby $x^0 = c t$ is interpreted as time. Further explanations on our notations and conventions can be found in the appendix.}
\begin{align}
\label{maxwell}
 \nabla_i B^i &= 0\ , & \nabla_0 B^i + \epsilon^{ijk}\partial_j E_k &= 0\ , \\
\label{maxwell2}
 \nabla_i D^i &= \rho\ , & \epsilon^{ijk} \partial_j H_k - \nabla_0 D^i &= j^i\ .
\end{align}
To account for possibly curvilinear coordinates we used the the covariant derivative in three dimensions, $\nabla_i$, with
\begin{equation}
\label{covder}
 \nabla_i A^i = (\partial_i + \Gamma^i_{i j}) A^j = \frac{1}{\sqrt{\gamma}} \partial_i(\sqrt{\gamma}A^i)\ ,
\end{equation}
where $\gamma$ is the determinant of the space metric $\gamma_{ij}$. Now a diffeomorphism to a virtual space called electromagnetic space is defined, which locally is implemented as a coordinate transformation $x^\mu \rightarrow \bar x^\mu$. Its effect is captured by re-writing the Maxwell equations in terms of the new, barred variables. More involved is the new relation among the fields $\bar{ \bm D}$, $\bar{ \bm B}$, $\bar{ \bm E}$ and $\bar{ \bm H}$, which in a generic coordinate system takes the form \cite{Landau2}
\begin{align}
\label{epsorig}
 \bar D^i &= \varepsilon_0 \frac{\bar \gamma^{ij}}{\sqrt{-\bar g_{00}}} \bar E_j - \frac{\bar g_{0j}}{\bar g_{00} c} \bar \epsilon^{jil} \bar H_l\ , &
 \bar B^i &= \mu_0 \frac{\bar \gamma^{ij}}{\sqrt{-\bar g_{00}}} \bar H_j + \frac{\bar g_{0j}}{\bar g_{00} c} \bar \epsilon^{jil} \bar E_l\ .
\end{align}
Here, $\bar g_{\mu\nu}$ are the components of the transformed spacetime metric, from which the transformed space metric follows as $\bar \gamma^{ij} = \bar g^{ij}$ (see Eq.\ \eqref{indmetric}). These relations resemble the constitutive relations of a special medium, but of course just describe the same vacuum solution as introduced above, re-written in complicated coordinates. To make use of the relations \eqref{epsorig} as medium parameters, the solutions $\bar{ \bm D}$, $\bar{ \bm B}$, $\bar{ \bm E}$ and $\bar{ \bm H}$ must be turned back into solutions in terms of the metric $g_{\mu\nu}$ in the coordinate system $x^\mu$, while keeping the form of the ``constitutive relations'' \eqref{epsorig} in terms of $\bar g_{\mu\nu}$. Since the Maxwell equations only depend on the determinant of the metric, but not on its specific components, this can be achieved by the simple rescaling \cite{Leonhardt:2006Nj,Bergamin:2008Pa}
\begin{align}
\label{scal1}
 \tilde{\bm E} &= \bar{\bm E}\ , & \tilde{\bm B} &= \frac{\sqrt{\bar \gamma}}{\sqrt{\gamma}} \bar{\bm B}\ , &
 \tilde{\bm D} &= \frac{\sqrt{\bar \gamma}}{\sqrt{\gamma}}  \bar{\bm D}\ , & \tilde{\bm H} &= \bar{\bm H}\ .
\end{align}
If  $\bar{ \bm D}$, $\bar{ \bm B}$, $\bar{ \bm E}$ and $\bar{ \bm H}$ are a solution of the Maxwell equations with metric $\bar g_{\mu\nu}$ and with ``constitutive relations'' \eqref{epsorig}, then $\tilde{ \bm D}$, $\tilde{ \bm B}$, $\tilde{ \bm E}$ and $\tilde{ \bm H}$ are a solution in terms of the coordinates $x^\mu$ with metric $g_{\mu\nu}$ and with constitutive relations
\begin{align}
\label{epstilde2}
 \tilde{ D}^i &= \varepsilon_0 \frac{\bar g^{ij}}{\sqrt{-\bar g_{00}}}  \frac{\sqrt{\bar \gamma}}{\sqrt{\gamma}} \tilde E_j - \frac{\bar g_{0j}}{\bar g_{00}c} \epsilon^{jil} \tilde{ H}_l\ , &
 \tilde B^i &= \mu_0 \frac{\bar g^{ij}}{\sqrt{-\bar g_{00}}} \frac{\sqrt{\bar \gamma}}{\sqrt{\gamma}} \tilde{ H}_j + \frac{\bar g_{0j}}{\bar g_{00}c} \epsilon^{jil} \tilde E_l\ .
\end{align}
In contrast to Eq.\ \eqref{epsorig}, which still describe electrodynamics in empty space, the constitutive relations \eqref{epstilde2} describe electrodynamics in a medium\footnote{In this sense, $\bar{g}_{\mu\nu}$ in \eqref{epstilde2} is sometimes called the ``optical'' metric, as opposed to the ``background'' metric $g_{\mu\nu}$.}. The basic idea of transformation optics is illustrated in Fig.\ \ref{fig:TOnotation}, which also summarizes our notation.

From the basic characteristics of transformation optics it follows that a negative index of refraction can appear at two different places, which motivates the first two questions to address:
\begin{enumerate}
 \item Is there any ``vacuum with negative index of refraction'' that could be mimicked by transformation optics? This question obviously is equivalent to the more general questions: Does there exist a formulation of electrodynamics where empty space has a negative index of refraction? Or: Does there exist any (usually improper) coordinate transformation such that the index of refraction from Eq.\ \eqref{epsorig} becomes negative? The answer is negative and in Sec.~\ref{se:improper} the basic derivation of this result is reviewed.
 \item Even though there exist no vacuum solutions with negative index of refraction, such a change of sign in permittivity and permeability can still originate from the second step of transformation optics, the re-interpretation of the solutions \cite{Leonhardt:2006Nj}. Thus we have to ask: Is such a change of sign allowed and if so, is it unique? The answer to the first part of the question is positive, as already shown in \cite{Leonhardt:2006Nj}. However, there exists no unique way to introduce a negative index of refraction in transformation optics. As will be shown in Sec.~\ref{se:negativerefr}, a careful derivation of the constitutive relations naturally leads to a formulation without negative index.
\end{enumerate}
Since the second step in transformation optics (the re-interpretation of the barred solution) is an \emph{ad-hoc} procedure one still might include negative refraction in transformation optics to extend its range of applicability. This leads to the third question:
\begin{enumerate}
\addtocounter{enumi}{2}
 \item Can a sign choice that leads to negative refraction be justified and do the ensuing concepts correctly describe the physical behavior of media with negative index of refraction? The answer is not definite. In Sec.~\ref{se:interface} two possible conventions that yield to negative refraction are assessed. It is shown that a certain scenario (originally presented in Ref.\ \cite{Bergamin:2008Pa}) can be motivated from the study of boundary conditions at interfaces between two different transformation media. Nonetheless, as shown in Sec.~\ref{se:lenses}, results from transformation optics making use of negative index of refraction and the associated multi-valued maps should still be used with utmost care. In particular, the transformation optics version of the Pendry-Veselago lens neither amplifies evanescent modes nor includes an imaging of the near field.
\end{enumerate}

\section{Maxwell's equations and {improper} coordinate transformations}
\label{se:improper}
Since the discussion of the first question mainly deals with the symmetries of electrodynamics in spacetime, it is most easily discussed in a relativistically covariant formulation. Following the ideas of Ref.\ \cite{Hehl:2003} we define a field strength tensor $F$ and an excitation tensor $\mathcal H$, which a priori are seen as independent quantities that are then related by the constitutive relation. In form language \cite{Hehl:2003,Lindell:DiffForms} the excitation tensor obeys the inhomogeneous Maxwell equation (Maxwell-Amp\`ere equation)
\begin{equation}
 \extd \mathcal H = \mathcal J ,
\end{equation}
which is seen to follow directly from the conservation of charge ($\extd \mathcal J = 0$). Here, $\mathcal J$ is a three form and consequently $\mathcal H$ is a two-form. Experimentally the charge is determined by means of a counting process which, as explained in detail in Ref.\ \cite{Hehl:2003}, chap.~B.1, does not require a spacetime metric (measure of space and time). Crucially, charge-counting is also observed to be independent of the orientation, so that $\mathcal J$ must be a \emph{twisted} three form. This implies that $\mathcal H$ is twisted as well, which means that it transforms under a generic change of frame $L_{\mu}{}^{\nu}$ as
\begin{equation}
 \bar{\mathcal H}_{\mu\nu} = \sgn(\det L_{\mu}{}^{\nu}) L_{\mu}{}^{\rho}L_{\nu}{}^{\sigma} \mathcal H_{\rho\sigma}\ ,
\end{equation}
where the additional sign $\sgn(\det L_{\mu}{}^{\nu})$ is called twist.

The field-strength tensor must be a closed two-form $\extd F = 0$, thus generating the homogeneous Maxwell equations (Maxwell-Faraday equations). In addition, $F$ can be combined with the current $\mathcal{J}$ to obtain the Lorentz force density $f_{\mu}$, which is a four-form carrying a covector value (Ref.\ \cite{Hehl:2003}, chap.~B.2 ):
\begin{equation}
 f_{\mu} = (e_{\mu}\din F)\wedge \mathcal J\ ,
\end{equation}
where $e_{\mu}$ is a component of the frame (base vector of the cotangent space). From observational data we know that $f_{\mu}$ must be a twisted form and thus $F$ is untwisted since already $\mathcal J$ is twisted.

Finally, $\mathcal H$ and $F$ are related by the constitutive relation. From the above discussion it follows immediately that in a linear relation,
\begin{equation}
\label{kappadef}
 \mathcal H = \kappa | F\ ,
\end{equation}
$\kappa$ must be a twisted $(2,2)$ tensor (a bi-vector valued two form). Therefore, the transformations of the effective medium parameters are fixed completely by the definition \eqref{kappadef}.

An important special case of Eq.\ \eqref{kappadef} is the relation between $\mathcal H$ and $F$ in empty space, $\mathcal H = \ast F$, where $\ast$ is the Hodge star operation. At this point it is important to notice that the Hodge star contains the totally anti-symmetric Levi-Civita symbol,
\begin{equation}
 (\ast F)_{\mu\nu} = \frac{1}{2} \epsilon_{\mu\nu}{}^{\rho\sigma} F_{\rho\sigma}\ ,
\end{equation}
which is a twisted tensor:
\begin{gather}
 \epsilon_{\mu\nu\rho\sigma} = \sqrt{-g} [\mu\nu\rho\sigma]\ , \qquad [0123] = 1\ ,\\
 \bar \epsilon_{\mu\nu\rho\sigma} = \sgn(\det L_{\mu}{}^{\nu}) L_{\mu}{}^{\alpha} L_{\nu}{}^{\beta} L_{\rho}{}^{\gamma} L_{\sigma}{}^{\delta} \epsilon_{\alpha\beta\gamma\delta}\ .
\end{gather}
The twist is necessary to ensure a unique definition of the tensor in the first line and leads to the shorter transformation rule
\begin{equation}
\label{simpletrans}
 \bar \epsilon_{\mu\nu\rho\sigma} = \frac{\sqrt{-\bar g}}{\sqrt{-g}} \epsilon_{\mu\nu\rho\sigma}\ .
\end{equation}
Therefore the relation $\mathcal H = \ast F$ indeed maps an untwisted tensor onto a twisted one. It is a widespread choice that none of the quantities in (say) $\bm D = \varepsilon_0 \bm E$ and $\bm B = \mu_0 \bm H$ carries a twist. Fortunately, it is possible manipulate \eqref{kappadef} so to accommodate for this popular convention. We shall now define a constitutive relation that maps an untwisted field excitation into an untwisted field strength. To ``untwist'' the excitation tensor it simply has to be mapped onto its Hodge dual. Since $\ast \ast \psi = - \psi$, for any two form $\psi$ in a four-dimensional Minkowski vector space, we define
\begin{equation}
 \mathcal G = - \ast \mathcal H
\end{equation}
and arrive at the constitutive relation
\begin{align}
\label{GFconstrel}
 \mathcal G &= \chi | F\ , & \chi = - \ast \kappa\ ,
\end{align}
which indeed leads us to the empty space relation $\mathcal G = F$. As is easily seen all quantities in this form of the constitutive relation are untwisted tensors. In the component formulation it is advantageous to raise the two indices of $\mathcal G$ by means of the metric, so that\footnote{Although \eqref{GFconstrel}$_{1}$ and \eqref{constindex} resemble the notation of Refs.~\cite{Post,Hehl:2003}, \eqref{GFconstrel}$_{2}$ indicates that different definitions are used here. Furthermore, in our derivation $\mathcal G$ is obtained from $\mathcal H$ by a Hodge star. This extra step is omitted in Refs.~\cite{Post,Hehl:2003}, where only pre-metric procedures are considered.}
\begin{equation}
 \mathcal G^{\mu\nu} = \frac{1}{2} \chi^{\mu\nu\rho\sigma} F_{\rho\sigma}.\label{constindex}
\end{equation}
Since $\chi$ transforms quartic under a generic change of frame it is obvious that it does not change sign under improper coordinate transformations. Consequently, $\bm D = \varepsilon_0 \bm E$ and $\bm B = \mu_0 \bm H$ cannot be transformed into a relation with formally negative index of refraction. Considering the permittivity, this is easily seen from the identification
\begin{equation}
 \varepsilon_R^{ij} = - \chi^{0i0j}\ .
\end{equation}
Under purely spatial transformations $\varepsilon_R^{ij}$ thus transforms in the same way as the spatial metric,
\begin{equation}
 \bar \varepsilon_R^{ij} = - (L^{-1})_i{}^k (L^{-1})_j{}^l \chi^{0k0l} = (L^{-1})_i{}^k (L^{-1})_j{}^l \varepsilon_R^{kl}\ ,
\end{equation}
and therefore its sign \emph{cannot} be changed.

A recent paper in theoretical electrodynamics (Ref. \cite{DaRocha:2010}) claims that twist factors could be ignored, thus invalidating many of our arguments. However,  Refs. \cite{Itin:2009ap , Post} clearly disprove this position, and expose unavoidable inconsistencies. In a different context, Refs. \cite{Mccall:2007prl , Mccall:2008meta} also point out that classical vacuum cannot refract negatively. Fortified by these observations, we now proceed to the re-interpretation step of transformation optics.

\section{Negative index of refraction in transformation optics}
\label{se:negativerefr}
The transformation properties found in the previous section just cover half of the program of transformation optics. In a second step the new solution found by the transformation, $\bar{\bm E}$, $\bar{\bm B}$, $\bar{\bm D}$ and $\bar{\bm H}$ has to be re-interpreted as a solution in the space with metric $g_{\mu\nu}$ instead of $\bar g_{\mu\nu}$. If the diffeomorphism $x^\mu \rightarrow \bar{x}^\mu$ changes the orientation, then obviously this re-interpretation $\tilde{x}^\mu=\bar{x}^\mu(x)$ has to change back to the original orientation. This step again can lead to additional signs and eventually to a negative index of refraction.

Since re-interpretation is an \emph{ad-hoc} manipulation rather than a mathematically strictly defined procedure, the exact equations to be manipulated have to be defined first. Equations \eqref{maxwell} and \eqref{maxwell2} represent a valid choice. Nonetheless, the covariance of Maxwell's equations plays a central role in what follows. Consequently, it is more appropriate to opt for the tensor index notation
\begin{align}
\label{faraday}
 \epsilon^{\mu\nu\rho\sigma}\partial_{\nu} F_{\rho\sigma}&=0\ , & & \text{Maxwell-Faraday equations;}\\
 \label{ampere}
 D_{\mu} \mathcal{G}^{\mu\nu} = \frac{1}{\sqrt{-g}} \partial_\mu (\sqrt{-g} \mathcal G^{\mu\nu})&= J^\nu\ , && \text{Maxwell-Amp\`ere equations;}
\end{align}
with the constitutive relation \eqref{GFconstrel}, whereby Eqs.\ \eqref{maxwell} and \eqref{maxwell2} are obtained from the identifications
\begin{align}
\label{Fident}
 E_i &= F_{0i}\ , & B^i &= -\frac{1}{2c} \epsilon^{ijk} F_{jk}\ ,\\
 \label{Hident}
 D^i &= - \epsilon_0 \sqrt{-g_{00}} \mathcal G^{0i}\ , & H_i &= - \frac{\epsilon_0c \sqrt{-g_{00}}}{2} \epsilon_{ijk} \mathcal G^{jk}\ .
\end{align}

In the rigorous treatment of Sec.~\ref{se:improper}, $\mathcal G$ is a untwisted tensor  and no sign ambiguities can appear in the Maxwell-Amp\`ere equations.  Thus, given the covariance of Eq.\ \eqref{ampere} under $x^\mu \rightarrow \bar{x}^\mu$, the identifications
\begin{align}
 \tilde{\mathcal G}^{\mu\nu} &= \frac{\sqrt{-\bar g}}{\sqrt{-g}} \bar{\mathcal G}^{\mu\nu}\ ,& {{\tilde{J}^{\nu}}} &{= \frac{\sqrt{-\bar g}}{\sqrt{-g}} \bar{J}^{\nu}}\ ,
\end{align}
convert the solution $\bar{\mathcal G}$ with spacetime $\bar g_{\mu\nu}$ back into a solution with spacetime $g_{\mu\nu}$.
Equation \eqref{faraday} (Maxwell-Faraday) includes the Levi-Civita tensor and thus might change sign. But due to relation \eqref{simpletrans} one can write
\begin{equation}
 \bar \epsilon^{\mu\nu\rho\sigma}\bar \partial_{\nu} \bar F_{\rho\sigma}= \frac{\sqrt{-g}}{\sqrt{-\bar g}} \epsilon^{\mu\nu\rho\sigma}\bar \partial_{\nu} \bar F_{\rho\sigma}= 0 \Rightarrow \epsilon^{\mu\nu\rho\sigma}\bar \partial_{\nu} \bar F_{\rho\sigma}= 0.
\end{equation}
Here, direct comparison with $\epsilon^{\mu\nu\rho\sigma} \partial_{\nu} \tilde F_{\rho\sigma}= 0$ implies the simple identification
\begin{equation}
 \bar F_{\mu\nu} = \tilde F_{\mu\nu},
\end{equation}
with no sign ambiguity\footnote{The same result is found in terms of Eqs.~\eqref{maxwell} and \eqref{maxwell2}. Again the strategy is to rewrite $\bar \epsilon^{ijk} = \sqrt{\gamma}/\sqrt{\bar \gamma}\epsilon^{ijk}$ to remove all sign ambiguities. Of course, one has to be careful -- the three dimensional Levi-Civita tensor also appears in the definition of the magnetic fields, which thus are twisted vector fields under spatial transformations.}. This indicates that the simple rescalings \eqref{scal1} and the constitutive relations \eqref{epstilde2} are generically valid. Hence, strictly speaking, \emph{there is no room for a negative index of refraction within transformation optics.}

At this point, one might be surprised to see that Refs.~\cite{Leonhardt:2006Nj, Bergamin:2008Pa} associate a negative index with orientation-reversing diffeomorphisms. This {is because these papers (more or less consciously) exploit the flexibility offered by the re-interpretation step. Additional signs are inserted when reverting to $\tilde{x}^\mu=\bar{x}^\mu(x)$, thus ``mimicking'' a negative refraction. For example in Ref.\ \cite{Leonhardt:2006Nj}, one must rescale the fields as
\begin{align}
\label{scal3}
 \tilde{\bm E} &= \bar \sigma \bar{\bm E}\ , & \tilde{\bm B} &= \frac{\sqrt{\bar \gamma}}{\sqrt{\gamma}} \bar{\bm B}\ , &
 \tilde{\bm D} &= \frac{\sqrt{\bar \gamma}}{\sqrt{\gamma}} \bar{\bm D}\ , & \tilde{\bm H} &= \bar \sigma \bar{\bm H}\ ,
\end{align}
with $\bar \sigma = \sgn(\det L_{i}{}^{j})$ (the sign of the \emph{spatial} transformation).  Crucially, this re-interpretation affects the constitutive relations \eqref{epstilde2}, which now become
\begin{align}
\label{epstilde3}
 \tilde{ D}^i &= \bar \sigma\frac{\bar g^{ij}}{\sqrt{-\bar g_{00}}}  \frac{\sqrt{\bar \gamma}}{\sqrt{\gamma}} \tilde E_j - \frac{\bar g_{0j}}{\bar g_{00}} \epsilon^{jil} \tilde{ H}_l\ , &
 \tilde B^i &= \bar \sigma\frac{\bar g^{ij}}{\sqrt{-\bar g_{00}}} \frac{\sqrt{\bar \gamma}}{\sqrt{\gamma}} \tilde{ H}_j + \frac{\bar g_{0j}}{\bar g_{00}} \epsilon^{jil} \tilde E_l\ .
\end{align}
As a consequence, according to Ref.~\cite{Leonhardt:2006Nj}, \emph{transformations that change the orientation of the coordinate system yield media with negative index.} It should be pointed out that such a scenario does not violate any basic requirements of transformation optics, even though the specific choice of signs may look somewhat arbitrary.
Ref. \cite{Bergamin:2008Pa} proposes a more consistent program, which also includes spacetime transformations. In this case, the rescaling concerns the Faraday tensor, which is re-defined as
\begin{equation}
 \tilde F_{\mu\nu} = \bar s \bar F_{\mu\nu}\ .
\end{equation}
with $\bar{s}=\sgn(\det L_{\mu}{}^{\nu})$. Here, one should notice that a map $\bar t = -t$ also yields a negative index of refraction\footnote{Historically, authors have linked negative refraction to improper coordinate transformations and to an extra sign in the cross product \cite{Leonhardt:2006Nj}. In the light of Sec.~\ref{se:improper}, however, it appears that the resulting negative index is more a matter of re-interpretation.}. In addition, the signs $\bar \sigma$ in Eq.\ \eqref{epstilde3} have to be replaced by $\bar s$ in this prescription.

In conclusion, a mathematically strict derivation of the constitutive relations of transformation optics suggests that a negative index of refraction should not be part of this program. Nonetheless, there exists a certain freedom of choice of signs and therefore transformation optics including negative index media can be conceived. The three prescriptions presented here are summarized in Table \ref{tab:summary}. The list is not exhaustive -- further possibilities to distribute the signs are available.
\begin{table}[t]
\begin{center}
\[
 \begin{array}{|l|ccc|}
  \hline&&&\\[-1.6ex]
  \mbox{Type} & \mbox{(A): \cite{Leonhardt:2006Nj,Leonhardt:2008Oe}} & \mbox{(B): \cite{Bergamin:2008Pa}} & \mbox{(C): Sec.~\ref{se:improper}}\\ \hline&&&\\[-1.6ex]
  \mbox{Spacetime Tensors} & \tilde F_{\mu\nu} = \bar\sigma F_{\mu\nu} & \tilde F_{\mu\nu} = \bar s F_{\mu\nu} & \tilde F_{\mu\nu} = F_{\mu\nu} \\
   & \tilde{\mathcal G}^{\mu\nu} = \frac{\sqrt{-\bar g}}{\sqrt{-g}} \bar{\mathcal G}^{\mu\nu} & \tilde{\mathcal G}^{\mu\nu} = \frac{\sqrt{-\bar g}}{\sqrt{-g}} \bar{\mathcal G}^{\mu\nu} & \tilde{\mathcal G}^{\mu\nu} = \frac{\sqrt{-\bar g}}{\sqrt{-g}} \bar{\mathcal G}^{\mu\nu} \\ \hline&&&\\[-1.6ex]
  \mbox{Polar Vectors} & \tilde{\bm E} =  \bar \sigma \bar{\bm E} & \tilde{\bm E} =  \bar s \bar{\bm E} & \tilde{\bm E} =  \bar{\bm E} \\
  & \tilde{\bm D} =  \frac{\sqrt{\bar \gamma}}{\sqrt{\gamma}} \bar{\bm D} & \tilde{\bm D} =  \frac{\sqrt{\bar \gamma}}{\sqrt{\gamma}} \bar{\bm D} & \tilde{\bm D} =  \frac{\sqrt{\bar \gamma}}{\sqrt{\gamma}} \bar{\bm D} \\\hline&&&\\[-1.6ex]
  \mbox{Axial Vectors} & \tilde{\bm B} = \frac{\sqrt{\bar \gamma}}{\sqrt{\gamma}} \bar{\bm B} & \tilde{\bm B} = \bar s \bar \sigma\frac{\sqrt{\bar \gamma}}{\sqrt{\gamma}} \bar{\bm B} & \tilde{\bm B} = \bar \sigma \frac{\sqrt{\bar \gamma}}{\sqrt{\gamma}} \bar{\bm B} \\
  & \tilde{\bm H} = \bar \sigma \bar{\bm H} & \tilde{\bm H} = \bar \sigma \bar{\bm H} & \tilde{\bm H} = \bar \sigma \bar{\bm H} \\ \hline&&&\\[-1.6ex]
  \mbox{Negative Index} & x^i\rightarrow \bar x^i &  x^\mu\rightarrow \bar x^\mu & \mbox{never} \\
  \mbox{of Refraction} & \mbox{changes orientation} & \mbox{changes orientation} & \\ \hline
 \end{array}
\]
%
  \end{center}
 \caption{Summary of the three discussed options to treat orientation changing transformations. $\bar \sigma$ takes value $-1$ if the spatial coordinate system changes orientation, $\bar s = -1$ if the spacetime coordinates change orientation.}
 \label{tab:summary}
\end{table}
\section{Boundary conditions and reflectionless interfaces}
\label{se:interface}
Though a strict application of the transformation rules must lead to the conclusion that media with a negative index of refraction cannot be covered by transformation optics, such a scenario can still be implemented thanks to the freedom in the interpretation process (from transformed vacuum to material). Keeping this in mind, we shall address the third question formulated in the introduction: which choice of signs, yielding to a negative index of refraction, should be selected and how realistic are the resulting negative refracting media? As far as the first part of the question is concerned we will not give a complete answer. We shall focus on the comparison of options (A) and (B) in Table \ref{tab:summary}.

To assess these two scenarios we consider an additional key ingredient, \emph{interfaces}. These play a crucial role in the theory of negative refracting media and in the particular application of perfect lenses. As customary, any interface -- \emph{even} between two transformation media -- can be discussed in terms of medium (tilde) fields. Nonetheless, when dealing with boundaries and transformation optics, it is preferable to relate the problem \emph{back} to the \emph{vacuum} solutions\footnote{Indeed, one should try and use the extra functionality offered by the transformation algorithm. In other words, one should exploit fully the program described in Fig.~\ref{fig:TOnotation}.}, as pointed out in Ref.~\cite{Bergamin:2009In}. Following this strategy, we shall use the interface conditions and compare options (A) and (B) of Table \ref{tab:summary}.

Let us consider a passive interface between a ``medium on the left'' (index $L$) and a ``medium on the right'' (index $R$) with standard dielectric boundary conditions
\begin{align}
\label{BC1}
 (\tilde{\bm D}_L - \tilde{\bm D}_R )\cdot \bm n &= 0\ , & (\tilde{\bm B}_L - \tilde{\bm B}_R )\cdot \bm n &= 0\ , \\
 \label{BC2}
 (\tilde{\bm E}_L - \tilde{\bm E}_R) \times \bm n &= 0\ , & (\tilde{\bm H}_L - \tilde{\bm H}_R) \times \bm n &= 0\ ,
\end{align}
where $\bm n$ is a vector normal to the interface. The key goal is to re-cast these four conditions by virtue of transformation optics. All medium solutions of Maxwell's equations, then, must be linked back to their vacuum counterparts (according to the process of Fig.~\ref{fig:TOnotation}). 

As a start, one sets all bi-anisotropic terms in the constitutive relations to zero\footnote{This has little consequences, since we are mainly interested in reflectionless interfaces. In terms of the transformation of space this implies that the map $x^\mu \rightarrow \bar x^\mu$ does not mix space $x^i$ and time $x^0$.}. Then, given the identifications \eqref{Fident} and \eqref{Hident}, with the transformation laws
\begin{align}
 \bar F_{\mu\nu} &= \frac{\partial x^{\rho}}{\partial \bar x^{\mu}}\frac{\partial x^{\sigma}}{\partial \bar x^{\nu}} F_{\rho\sigma}\ , & \bar{\mathcal G}^{\mu\nu} &= \frac{\partial \bar x^{\mu}}{\partial x^{\rho}}\frac{\partial \bar x^{\nu}}{\partial  x^{\sigma}} \mathcal G^{\rho\sigma}\ ,
\end{align}
one obtains that the space vectors must transform as
\begin{align}
\label{barEB}
 \bar E_i &= \frac{\partial x^0}{\partial \bar x^0} \frac{\partial x^j}{\partial \bar x^i}  E_j\ , & \bar B^i &= \frac{\partial \bar x^i}{\partial x^j} B^j\ , &
 \bar D^i &=  \frac{\sqrt{-\bar g_{00}}}{\sqrt{-g_{00}}} \frac{\partial \bar x^0}{\partial x^0} \frac{\partial \bar x^i}{\partial x^j} D^j\ , & \bar H_i &=  \frac{\sqrt{-\bar g_{00}}}{\sqrt{-g_{00}}} \frac{\partial  x^j}{\partial \bar x^i} H_j\ .
\end{align}
At this point, the ``negative index schemes'' of Table \ref{tab:summary} enter the calculations. The medium solutions corresponding to $\bm D$ and $\bm H$ are the same both for (A) and for (B):
\begin{align}
 \label{tildeDH}
 \tilde D^i\left(\tilde x = \bar x(x)\right) &=   \frac{\sqrt{-\bar g}}{\sqrt{-g}} \frac{\partial \bar x^0}{\partial x^0} \frac{\partial \bar x^i}{\partial x^j} D^j\ , &
 \tilde{H}_i\left(\tilde x = \bar x(x)\right) &=  \bar \sigma \frac{\partial x^0}{\partial \bar x^0} \frac{\partial x^j}{\partial \bar x^i} H_j(x)\ ,
\end{align}
where one should recall \eqref{detrel}. The relation for $\bm D$ can further be simplified by using
\begin{align}
 \bar g_{\mu\nu} &= \frac{\partial x^{\rho}}{\partial \bar x^{\mu}}\frac{\partial x^{\sigma}}{\partial \bar x^{\nu}} g_{\rho\sigma} &&\Longrightarrow & \sqrt{-\bar g} &= \sqrt{\left\lvert \frac{\partial x^\mu}{\partial \bar x^\nu}\right\lvert^2} \sqrt{-g} = \bar s \frac{\partial x^0}{\partial \bar x^0} \left\lvert \frac{\partial x^k}{\partial \bar x^l}\right\lvert  \sqrt{-g}\ . 
\end{align}
Analogously, the medium solutions for $\bm E$ and $\bm B$ are obtained, but these results depend on the chosen scenario ((A) or (B)). In summary:
\begin{align}
 \label{tildeDH2}
 && \tilde D^i\left(\tilde x = \bar x(x)\right) &= \alpha \bar s \left\lvert \frac{\partial x^k}{\partial \bar x^l}\right\lvert \frac{\partial \bar x^i}{\partial x^j} D^j(x) \ , &
 \tilde{H}_i\left(\tilde x = \bar x(x)\right) &= \alpha \bar s \frac{\partial x^0}{\partial \bar x^0} \frac{\partial x^j}{\partial \bar x^i} H_j(x)\ ,\\
\label{tildeEBA}
 (A): && \tilde{E}_i\left(\tilde x = \bar x(x)\right) &= \alpha \bar \sigma \frac{\partial x^0}{\partial \bar x^0} \frac{\partial x^j}{\partial \bar x^i} E_j(x)\ , & \tilde B^i\left(\tilde x = \bar x(x)\right) &= \alpha \bar \sigma \left\lvert \frac{\partial x^k}{\partial \bar x^l}\right\lvert \frac{\partial \bar x^i}{\partial x^j} B^j(x) \ , \\
 \label{tildeEBB}
 (B): && \tilde{E}_i\left(\tilde x = \bar x(x)\right) &= \alpha \bar s \frac{\partial x^0}{\partial \bar x^0} \frac{\partial x^j}{\partial \bar x^i} E_j(x)\ , & \tilde B^i\left(\tilde x = \bar x(x)\right) &= \alpha \bar s \left\lvert \frac{\partial x^k}{\partial \bar x^l}\right\lvert \frac{\partial \bar x^i}{\partial x^j} B^j(x) \ .
\end{align}
In these equations we have introduced a new parameter $\alpha$, which just represents the fact that shifting \emph{all} fields by a constant does not change the constitutive relations. Without loss of generality one can assume $\alpha = \pm 1$ and moreover $\alpha \equiv 1$ in the case of a trivial map $\bar x^\mu \equiv x^{\mu}$ \cite{Bergamin:2009In}. Eqs.\ \eqref{tildeDH2}--\eqref{tildeEBB} allow to re-express the boundary conditions \eqref{BC1} and \eqref{BC2} in terms of the vacuum solutions (as required). To facilitate the use of component notation we can assume an adapted coordinate system in laboratory space, such that the direction normal to the interface is labeled by the coordinate $\tilde{\bm x}_\perp = (0,0,\tilde x^\perp)$, while the directions parallel to the interface have coordinates $\tilde{\bm x}_\parallel = (\tilde x^A,0)$, whereby the index $A$ takes values $1,2$. In this notation the condition \eqref{BC1} becomes
\begin{align}
\label{newBC1}
 \alpha_L \bar s_L \left\lvert \frac{\partial x^k}{\partial \bar x_L^l}\right\lvert \frac{\partial \bar x_L^\perp}{\partial x^j} D_L^j &= \alpha_R \bar s_R \left\lvert \frac{\partial x^k}{\partial \bar x_R^l}\right\lvert \frac{\partial \bar x_R^\perp}{\partial x^j} D_R^j\ ,  & \alpha_L \bar s_L \frac{\partial x^0}{\partial \bar x_L^0} \frac{\partial x^j}{\partial \bar x_L^i} H^L_j  & = \alpha_R \bar s_L \frac{\partial x^0}{\partial \bar x_R^0} \frac{\partial x^j}{\partial \bar x_R^i} H^R_j\ ,
\end{align}
while \eqref{BC2} splits in two cases:
\begin{align}
\label{newBC2.1}
 (A)&& \alpha_L \bar s_L \frac{\partial x^0}{\partial \bar x_L^0} \frac{\partial x^j}{\partial \bar x_L^A} E^L_j &= \alpha_R \bar s_R \frac{\partial x^0}{\partial \bar x_R^0} \frac{\partial x^j}{\partial \bar x_R^A} E^R_j & \alpha_L \bar s_L \frac{\partial x^0}{\partial \bar x_L^0} \frac{\partial x^j}{\partial \bar x_L^i} H^L_j  & = \alpha_R \bar s_L \frac{\partial x^0}{\partial \bar x_R^0} \frac{\partial x^j}{\partial \bar x_R^i} H^R_j \\
 \label{newBC2.2}
 (B)&& \alpha_L \bar \sigma_L \frac{\partial x^0}{\partial \bar x_L^0} \frac{\partial x^j}{\partial \bar x_L^A} E^L_j &= \alpha_R \bar \sigma_R \frac{\partial x^0}{\partial \bar x_R^0} \frac{\partial x^j}{\partial \bar x_R^A} E^R_j & \alpha_L \bar \sigma_L \frac{\partial x^0}{\partial \bar x_L^0} \frac{\partial x^j}{\partial \bar x_L^i} H^L_j  & = \alpha_R \bar \sigma_L \frac{\partial x^0}{\partial \bar x_R^0} \frac{\partial x^j}{\partial \bar x_R^i} H^R_j
\end{align}
These steps to re-cast the boundary conditions might look like a mathematical exercise, but in fact they contain very important information. The idea of transformation optics is to design media that ``mimic a different space'' or ``mimic a transformation of coordinates'' (Refs.\ \cite{Leonhardt:2006Nj,Leonhardt:2008Oe,Bergamin:2008Pa}). This interpretation immediately applies to the medium on the left as well as to the medium on the right. But is the interface \emph{also} a region of mimicking? Fortunately, Eqs.~\eqref{newBC1}--\eqref{newBC2.2} allow to answer this question \cite{Bergamin:2009In}. The interface is a region of mimicking if the \emph{vacuum} solutions are continuous at the interface, so that
\begin{align}
\label{contvac}
 \bm E_L &= \bm E_R\ , & \bm B_L &= \bm B_R\ , & \bm D_L &= \bm D_R\ , & \bm H_L &= \bm H_R\ , &&\text{at the interface,}
\end{align}
which in particular implies the absence of reflections in the medium solutions.
The combination of \eqref{newBC1}--\eqref{newBC2.2} with \eqref{contvac} singles out all interfaces which fit into the picture of transformation optics. \emph{Firstly}, the conditions \eqref{newBC1} yield 
\begin{align}
\label{parallelI}
 \alpha_L \bar s_L \frac{\partial x^0}{\partial \bar x_L^0} \frac{\partial x^j}{\partial \bar x_L^A} &= \alpha_R \bar s_R \frac{\partial x^0}{\partial \bar x_R^0} \frac{\partial x^j}{\partial \bar x_R^A}\ , &
 \alpha_L \bar s_L \left\lvert \frac{\partial x^k}{\partial \bar x_L^l}\right\lvert \frac{\partial \bar x_L^\perp}{\partial x^j} &= \alpha_R \bar s_R \left\lvert \frac{\partial x^k}{\partial \bar x_R^l}\right\lvert \frac{\partial \bar x_R^\perp}{\partial x^j}\ .
\end{align}
and are satisfied if the two transformations obey \cite{Bergamin:2009In}
\begin{align}
\label{simple}
 \frac{\partial \bar x^A_L}{\partial \bar x^B_R} &= \delta^A_B\ , & \frac{\partial \bar x^0_L}{\partial \bar x^0_R} &= 1\ , & \alpha_L &= \bar s_L\ , & \alpha_R &= \bar s_R \ .
\end{align}
Here, $\partial \bar x_L^\perp/\partial \bar x_R^\perp$ remains unrestricted. In particular, $\partial \bar x_L^\perp/\partial \bar x_R^\perp < 0$ is permitted and a spatial inversion in the normal direction implies negative refraction. Also, the transformed coordinates parallel to the interface must agree on both sides, while the transformation in the orthogonal direction is continuous, but not necessarily differentiable at the interface. Furthermore, the time coordinates must agree on both sides. \emph{Secondly}, the restrictions on $\tilde{\bm E}$ and $\tilde{\bm B}$ replicate the conditions \eqref{parallelI} for option (B), while a \emph{new} set is derived for option (A). Thus, from the point of view of interfaces, scenario (B) appears as the most effective (and economic) choice.\footnote{Since time reversal is not permitted by Eq.\ \eqref{contvac} scenario (A) formally yields the same set of allowed interfaces.} 

In summary, it was found that certain interfaces between two transformation media with a different sign in the index of refraction may be interpreted as mimicking a transformation of space. Technically, interfaces that permit such an interpretation must be reflectionless and revert the spatial direction normal to the interface, but are not allowed to invert the time direction. In addition, it was shown that the scheme (B) is preferable, if one is to design (e.g.) a negative refracting lens.

\section{Perfect lenses and evanescent modes}
\label{se:lenses}
If negative index of refraction can be made part of transformation optics, how good are the outcomes of this interpretation? To our knowledge, the perfect lens \cite{Pendry:Pl2000} is  the only device where negative refraction was obtained via transformation optics \cite{Leonhardt:2006Nj}. Thus, we restrict to this example here. A flat lens is associated with the map \cite{Leonhardt:2006Nj}
\begin{equation}
\label{lensmap}
  z = \begin{cases}
           \bar z\ , &  \bar z<0\ ; \\
	   -{a} \bar z\ , & 0< \bar z<D\ ; \\
	   \bar z-({a}+1) D\ , &  \bar z>D\ .
          \end{cases}
\end{equation}
as demonstrated in Fig. \ref{fig:lensmap} (one must choose $a>0$). 
\begin{figure}[t]
 \centering
 \includegraphics[width=0.6\linewidth]{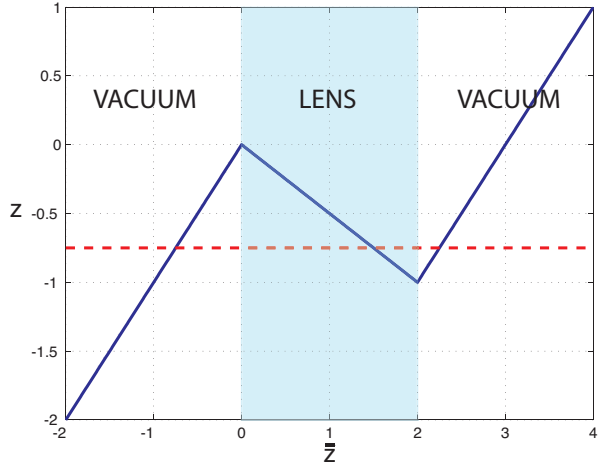}
\caption{Diagram illustrating the ``folding'' of space performed by \eqref{lensmap}. Here, the slab thickness $D$ is $2$ and the parameter $a=0.5$ (arbitrary units).  The dashed red line illustrates how three points along the $\bar{z}$ axis can be mapped to the same position $z$.}
 \label{fig:lensmap}
\end{figure}
As is easily seen, any point $-{a}D< z < 0$ is mapped on three different points in the virtual electromagnetic space and -- upon re-interpretation -- on three different points in laboratory space, whereby in the region $0<\tilde z<{a}D$ a medium with negative index of refraction emerges. This triple valued map was associated with perfect imaging. The key argument was that any solution of the Maxwell equations in the region $-{a}D< -z < 0$ was reproduced exactly inside the lens at $\tilde z = z/{a}$ and on the other side of the lens at $\tilde z = ({a}+1)D - z$.
\begin{figure}[t]
 \centering
 \includegraphics[width=\linewidth,bb=0 0 965 348]{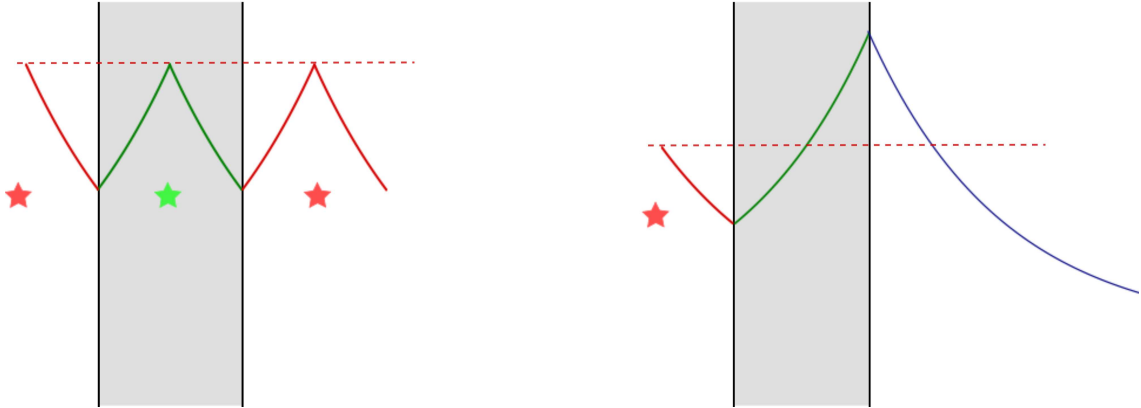}
\caption{Schematic view of evanescent modes in the the transformation optics lens (left hand side) and the Pendry-Veselago lens (right hand side). While the latter images by amplification of evanescent modes \cite{Pendry:Pl2000,Ramakrishna:2005Rp}, the former triples the sources. Once again, the dashed red line helps to follow the image formation process.}
 \label{fig:lens comparison}
\end{figure}

Since negative index of refraction within transformation optics is an effect of the choice of signs rather than an inherent characteristic, one should have a careful look at the lens proposed by the map \eqref{lensmap}. The following three conclusions are immediate:
\begin{enumerate}
 \item Due to causality, the transformation optics lens is strictly limited to stationary situations. It is well known that transformation devices can get in conflict with causality, but mostly this can be resolved by a limitation to a rather narrow bandwidth. The scenario envisaged here -- however -- is more pathological. The folding of space by means of the map \eqref{lensmap} limits the system to strictly stationary situations, simply because any change in the electromagnetic fields at the source point causes an \emph{instantaneous} change of the mirror image inside the lens and the image behind the lens.
 \item The transformation designed lens cannot image a source, but rather \emph{triples} it. Indeed, a situation with a source at the source point, but empty mirror image and image point, is not covered by transformation optics. Instead, a source automatically creates a mirror source (sink) inside the lens and a second source behind the lens (see Fig.\ \ref{fig:lens comparison}).
 \item Consequently, within transformation optics no enhancement of the evanescent waves takes place, which is the working principle of the Pendry-Veselago lens \cite{Pendry:Pl2000,Ramakrishna:2005Rp}. As can be seen from Fig.\ \ref{fig:lens comparison}, all evanescent waves in the transformation designed lens are easily explained as the evanescent modes generated by one of the three sources. There is no need for an amplification of such modes.
\end{enumerate}

\section{Conclusions}
\label{se:conclusion}
In this paper we have reviewed the status of a negative index of refraction within transformation optics. It was shown that negative refraction does not emerge from improper coordinate transformations, as ensured by the formalism of twisted tensors. Beside this fact, it was argued that negative index media could \emph{still} be included within transformation optics. The key point, in this respect, was to exploit the sign flexibility offered by re-interpretation algorithm. This justified the ``mimicking'' schemes of Refs.~\cite{Leonhardt:2006Nj, Bergamin:2008Pa}, which were subsequently compared and assessed -- revealing that the latter program is preferable.}

The most important application of a negative index, the perfect lens, was reviewed. Most importantly it was found that transformation designed lenses do not amplify the evanescent modes and are unable to image a source. This observation might impact recent ideas on perfect imaging -- with negative refraction or without (as in Refs.~\cite{Leonhardt:2009Pi,Leonhardt:2009Td}).

As a general conclusion it should be stressed that transformation designed imaging devices should be used with utmost care, in particular if they include negative index of refraction. In many cases, the analysis is essentially restricted to stationary situations where sources are not imaged, but rather duplicated. These findings should not come as a surprise -- negative refraction alone is not sufficient for near field imaging. Only the enhancement of evanescent waves associated to media with $\epsilon = \mu = -1$ enables this process (Ref.~\cite{Noginov:2009}, chap.~1).

\subsection*{Acknowledgment{s}}
The authors would like to thank S.~Tretyakov, C.~Simovski, I.~Nefedov, P.~Alitalo, M.W.~McCall and P.~Kinsler for stimulating discussions. This project was supported by the Academy of Finland, project no.\ 124204 and by the EPSRC, grant no. EP/E031463/1.

\appendix
\section{Covariant formulation}
In this Appendix we present our notations and conventions regarding the covariant formulation of the Maxwell equations on a generic (not necessarily flat) manifold and written in general coordinates. For a detailed introduction to the topic we refer to the relevant literature, e.g.\ \cite{Hehl:2003,Landau2,Post}.

Greek indices $\mu, \nu, \rho, \ldots$ are spacetime indices and run from 0 to 3, Latin indices $i,j,k,\ldots$ space indices with values from 1 to 3. Furthermore an adapted coordinate system is used at the interface, such that $(x^i) = (x^A,x^\perp)$, where $x^A$ are the directions parallel to the interface, while $x^\perp$ is perpendicular. Therefore capital Latin indices take values 1,2.
 
For the metric we use the ``mostly plus'' convention, so the standard flat metric is $g_{\mu\nu} = \mbox{diag}(-1,1,1,1)$. Time is always interpreted as the zero-component of $x^{\mu}$, $x^0 = c t$. With this identification an induced space metric can be obtained as \cite{Landau2}
\begin{equation}
\label{indmetric}
 \gamma^{ij} = g^{ij}\ , \qquad \gamma_{ij} = g_{lk} - \frac{g_{0i}g_{0j}}{g_{00}}\ , \qquad \gamma^{ij} \gamma_{jk} = \delta^i_k\ ,
\end{equation}
where $\delta^i_k$ is the Kronecker symbol.
This implies as relation between the determinant of the spacetime metric, $g$, and the one of the space metric, $\gamma$,
\begin{equation}
\label{detrel}
 -g = -g_{00} \gamma\ .
\end{equation}
The four dimensional Levi-Civita tensor is defined as
\begin{align}
 \epsilon_{\mu\nu\rho\sigma} &= \sqrt{-g}[\mu\nu\rho\sigma]\ , & \epsilon^{\mu\nu\rho\sigma} &= - \frac1{\sqrt{-g}}[\mu\nu\rho\sigma]\ ,
\end{align}
with $[0123] = 1$. The relation between the four-dimensional Levi-Civita tensors in two different spacetimes can be written as
\begin{equation}
 \epsilon_{\mu\nu\rho\sigma} = \frac{\sqrt{-g}}{\sqrt{-\bar g}} \bar \epsilon_{\mu\nu\rho\sigma}\ ,
\end{equation}
From Eq.\ \eqref{detrel} the reduction of the four dimensional to the three dimensional tensor follows as
\begin{equation}
 \epsilon_{0ijk} = \sqrt{-g_{00}} \epsilon_{ijk}\ , \qquad \epsilon^{0ijk} = - \frac1{\sqrt{-g_{00}}} \epsilon^{ijk}\ .
\end{equation}
If the Levi-Civita tensor is consistently treated as a twisted tensor these equations are invariant under improper transformations without any sign ambiguities.

The formulation of the Maxwell equations in spacetime starts from the definition of the field strength and excitation tensors, which both are two forms in four dimensional spacetime
\begin{align}
 F &= \frac{1}{2} F_{\mu\nu} \extd x^\mu\wedge \extd x^\nu\ , & \mathcal H &= \frac{1}{2} \mathcal H_{\mu\nu} \extd x^\mu\wedge \extd x^\nu\ .
\end{align}
These obey the Maxwell equations
\begin{align}
 \extd F &= 0\ , & \extd \mathcal H = \mathcal J\ ,
\end{align}
where $\extd$ indicates the exterior derivative, which maps $p$ forms onto $p+1$ forms. Instead of the excitation tensor $\mathcal H$ it is often advantageous to use its Hodge dual $\ast \mathcal H$. The Hodge star operation $\ast$ maps a $p$ form in $n$ dimensions onto a $(p-n)$ form, thus in four dimensional Minkowski space two forms are mapped onto two forms according to the rule
\begin{equation}
 \ast \mathcal H = \frac{1}{4} \mathcal H_{\lambda \tau} g^{\lambda \mu} g^{\tau\nu} \epsilon_{\mu\nu\rho\sigma} \extd x^\rho\wedge \extd x^\sigma = \frac{1}{4} \mathcal H^{\mu\nu} \epsilon_{\mu\nu\rho\sigma} \extd x^\rho\wedge \extd x^\sigma\ .
\end{equation}
The Hodge dual $\mathcal G = - \ast \mathcal H$ obeys the Maxwell equation
\begin{equation}
 \delta \mathcal G = J\ , \qquad J = - \ast \mathcal J\ .
\end{equation}
Here, $\delta = \ast \extd \ast$ is the coderivative and maps $p$ forms onto $p-1$ forms. Accordingly $J$ is a one form (a vector) in agreement with our standard interpretation of charges and currents in electrodynamics. In component notation the Maxwell equations can be displayed as
\begin{align}
\label{EOMcomp}
 \epsilon^{\mu\nu\rho\sigma} \partial_\nu F_{\rho \sigma} &= 0\ , & D_\nu \mathcal G^{\mu \nu} &= - J^\mu\ .
\end{align}
The Maxwell equations depend on the metric $g_{\mu\nu}$ through the covariant derivative $D_\mu$. Since
\begin{equation}
\label{covderII}
 D_\nu \mathcal G^{\mu \nu} = (\partial_\mu + \Gamma^{\nu}_{\nu \rho}) \mathcal G^{\mu \rho} = \frac{1}{\sqrt{-g}} \partial_{\nu} (\sqrt{-g} \mathcal G^{\mu\nu})
\end{equation}
it is seen that the Maxwell equations just depend on the determinant of the metric, but not on its individual components.

The space vectors $\bm E$ and $\bm B$ are found as components of the field strength tensor $F_{\mu\nu}$, while $\bm D$ and $\bm H$ become part of the excitation tensor $\mathcal G^{\mu\nu}$, with the identifications
\begin{align}
 [F_{\mu\nu}]&= \begin{pmatrix}
                  0 & E_1 & E_2 & E_3\\-E_1 & 0 & -c B^3 & c B^2 \\
		  -E_2 & c B^3 & 0 & -c B^1 \\
		  -E_3 & -c B^2 & c B^1 & 0
		  \end{pmatrix}\ , & [\mathcal G^{\mu\nu}]&= \frac{1}{\varepsilon_0 \sqrt{g_{00}}}\begin{pmatrix}
                  0 & -D^1 & -D^2 & -D^3\\ D^1 & 0 & - \frac{H_3}{c} & \frac{H_2}{c} \\
		  D^2 & \frac{H^3}{c} & 0 & -\frac{H^1}{c} \\
		  D^3 & -\frac{H_2}{c} & \frac{H_1}{c} & 0
		  \end{pmatrix}\ .
\end{align}
Finally, electric charge and current are combined into the four-current $J^{\mu} =(\sqrt{g_{00}} \varepsilon_0)^{-1} (\rho, j^i/c)$.

\bibliographystyle{fullsort}
\bibliography{bibliomaster}
\end{document}